\documentclass[envcountsame]{llncs}
\usepackage[llncs]{sylvie}
\usepackage{gastex}

\newcommand{\stc}{\operatorname{sc}}

\newcommand{\cyc}{{\rm{cyc}}}
\newcommand{\lquo}{\backslash}
\newcommand{\del}{\rightsquigarrow}

\title{A General Approach to State Complexity of Operations: Formalization and Limitations\thanks
{This work was supported by the Natural Sciences and Engineering Research Council of Canada grant No.~OGP0000871.}}
\author{Sylvie Davies}
\institute{Department of Pure Mathematics\\
University of Waterloo, Waterloo, Ontario, Canada\\
\email{sldavies@uwaterloo.ca}}

\begin{document}
\maketitle
\begin{abstract}
The state complexity of the result of a regular operation is often positively correlated with the number of distinct transformations induced by letters in the minimal deterministic finite automaton of the input languages. That is, more transformations in the inputs means higher state complexity in the output. When this correlation holds, the state complexity of a unary operation can be maximized using languages in which there is one letter corresponding to each possible transformation; for operations of higher arity, we can use $m$-tuples of languages in which there is one letter corresponding to each possible $m$-tuple of transformations.
In this way, a small set of languages can be used as witnesses for many common regular operations, eliminating the need to search for witnesses -- though at the expense of using very large alphabets.
We formalize this approach and examine its limitations. We define a class of ``uniform'' operations for which this approach works; the class is closed under composition and includes common operations such as star, concatenation, reversal, union, and complement.
Our main result is that the worst-case state complexity of a uniform operation can be determined by considering a finite set of witnesses, and this set depends only on the arity of the operation and the state complexities of the inputs. 

\end{abstract}

\section{Introduction}
Given a regular operation, how do we determine its (deterministic) state complexity? 
There is probably no universal method for solving these problems. 
Nonetheless, for many operations there is a common approach we can take. 
While this approach is not new, it does not seem to be universally known to researchers.
The key idea first appeared in a 1978 paper of Sakoda and Sipser~\cite{SaSi78}, but it has seldom been used in the context of state complexity. 
In cases where it was used, authors typically did not acknowledge the full power and generality  of the approach. This paper attempts to give a formal, general account of the approach and its uses in state complexity.

We will refer to the approach in question as the ``one letter per action'' (OLPA) approach.
We give an informal description of the OLPA approach below.
The root of the approach is to think about regular operations in terms of how they affect deterministic finite automata (DFAs). 
Typically, a regular operation takes some DFAs as input and modifies or combines them in some way to produce a new DFA. 
Many of these DFA constructions have the following properties: 
\bi
\item
If a new letter is added to the input DFAs, then the state complexity of the output DFA will not decrease. 
\item
If a new letter is added to the input DFAs, and in each DFA this letter acts the same as an existing letter, then the state complexity of the output DFA will stay the same.
\ei
If we know an operation has these two properties, this suggests a way to maximize the state complexity of the operation: keep adding new letters to the input DFAs, until the point where we have letters corresponding to all possible actions across all the input DFAs. 
The first property ensures that adding letters can only increase or maintain the state complexity of the output, while the second property ensures that the state complexity of the output will eventually reach a maximum.
In this way, we can obtain witnesses for the worst-case state complexity of the operation.

\begin{figure}[ht]
\unitlength 8.5pt
\begin{center}
\begin{picture}(8,4)(0,1)
\gasset{Nh=2,Nw=2,Nmr=1.25,ELdist=0.4,loopdiam=1.5}
\node(0)(0,2){$1$}
\node(1)(8,2){$2$}

\drawloop(0){$a_{11},a_{12}$}
\drawloop(1){$a_{12},a_{22}$}
\drawedge[curvedepth=1,ELdist=.4](0,1){$a_{21},a_{22}$}  
\drawedge[curvedepth=1,ELdist=.4](1,0){$a_{11},a_{21}$}  

\end{picture}
\end{center}
\caption{Two-state ``one letter per action'' DFA, with initial and final states unspecified.}
\label{fig:olpa}
\end{figure}
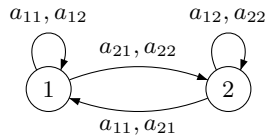

Figure \ref{fig:olpa} shows the result of applying this construction to a two-state DFA with unspecified initial and final states, to be used as input for a unary operation. The DFA has one letter for each of the four functions from $\{1,2\}$ to itself. To illustrate the construction for a three-state DFA, we would need $3^3 = 27$ letters!

\vspace{-0.5cm}
Since the state complexity of the output is not affected only by the actions of letters in the inputs, but also the initial states and final state sets of the inputs, we construct these witnesses for \e{each configuration} of initial and final states.
Then for each configuration, it remains to solve the combinatorial problem of counting reachable and pairwise distinguishable states in the output DFA, under the assumption that every possible action in the input DFAs is available.

Of course, these combinatorial problems can sometimes be quite hard, and this method produces witnesses over extremely large alphabets. 
Nonetheless, we believe the OLPA approach is useful to know for several reasons. 
For one, it gives a way to compute the exact worst-case state complexity of certain operations for inputs with small state complexity. 
Naively, computing this value would require checking all possible input DFAs under the state complexity threshold. 
However, if our operation has the aforementioned two properties, we can just check the small set of witness DFAs previously described.
This check is quite slow because of the large alphabets of the witnesses, but the computation is often feasible for a handful of small values, and often these small values are enough to start making conjectures about the general behaviour of the state complexity function.

Second, using ``OLPA witnesses'' with one letter for each action of the input DFAs can simplify proofs and make the ideas behind them more clear. State complexity proofs using witnesses over optimal or near-optimal alphabets are often rather technical. Consider the following proof that for $n \ge 2$, the state complexity of the reversal operation is $2^n$:

\e{Let $\cD = (Q,\Sig,\delta,q_0,F)$ be a DFA with $n$ states. 
Suppose for each function $t \co Q \ra Q$, there is a letter $a_t \in \Sig$ such that $\delta(q,a_t) = t(q)$. 
Let $\cR = (Q,\Sig,\Delta,F,\{q_0\})$ be an NFA for the reverse of $\cL(\cD)$, where $\Delta(q,a_t) = \{ p \in Q : \delta(p,a_t) = q\}$. 
For each $S \subseteq Q$, let $s$ be a function that maps $S$ into $F$ and $Q \setminus S$ into $Q \setminus F$. 
Then $\Delta(F,a_s) = \{ p \in Q : \delta(p,a_s) \in F \} = S$, so all $2^n$ subsets of $Q$ are reachable.
Distinct sets $S,T \subseteq Q$ are distinguished as follows: choose an element $q$ that (without loss of generality) is in $S$ but not $T$, and choose a function $s$ that maps $q_0$ to $q$ and $Q \setminus \{q_0\}$ into $Q \setminus \{q\}$.
Then $\Delta(S,a_s)$ contains the final state $q_0$ and $\Delta(T,a_s)$ does not, so the sets are distinguished.} \qed

Because we are free to choose letters that do exactly what we want, this proof is very simple.
It is also rather illuminating, since we can immediately extract a sufficient condition for attaining the worst-case state complexity from the proof: it suffices that there are letters which induce, for each $S \subseteq Q$, a function that maps $S$ into $F$ and $Q \setminus S$ outside of $F$, and for each $q \in Q$, a function that maps $q_0$ to $q$ and no other elements to $q$.
Furthermore, if one notices that letters can be replaced by words throughout the proof
(that is, the hypothesis
``there is a letter $a_t \in \Sig$ such that $\delta(q,a_t) = t(q)$''
can be replaced by
``there is a word $w_t \in \Sig^*$ such that $\delta(q,w_t) = t(q)$''),
one recovers the result of Salomaa, Wood and Yu~\cite{SWY04} that the state complexity is maximized if the transition monoid of $\cD$ contains all functions from $Q$ to itself (in addition to strengthening the aforementioned sufficient condition from letters to words).

For contrast, consider the proofs given by Jir\'askov\'a and \v{S}ebej~\cite{JiSe12} that the worst-case state complexity can be attained over ternary and binary alphabets.
The ternary proof is about as short as the proof above, but somewhat more terse, asking the reader to compute transitions under the word $bc^{i_2-i_1-1}a^{i_1}$. It also does not offer the same insight into general conditions for attaining the worst-case complexity.
Meanwhile, the binary proof is long and involves a complicated multi-case induction argument.
Of course, it is useful and desirable to have proofs that use witnesses with small alphabets, and we do not bring up these proofs to criticize them or suggest they should be replaced. 
We just believe that proofs using the OLPA approach can sometimes be simpler and more illuminating.

Our third reason for studying the OLPA approach is that we believe it could lead to a better general understanding of state complexity of operations, and the conditions that lead to maximal blow-ups in complexity. 
The fact that this approach exists and applies to many of the operations studied in the state complexity literature perhaps suggests something about the nature of state complexity. 
Suppose we think of each letter in an alphabet as an ``instruction'', and a sequence of instructions as a ``program''; then a language of state complexity $n$ can be viewed as a collection of programs for a ``computer'' with $n$ possible memory states.
It seems many operations attain their worst-case state complexity when provided with languages that have an optimal ``economy of description'' for programs -- one instruction (letter) for each possible program (action).
Is the key to finding witnesses over small alphabets to maximize this ``economy of description'' with respect to the alphabet size? 
Can the effectiveness of Brzozowski's ``universal witnesses''~\cite{Brz13}, which have small alphabets and maximize the complexity of several operations simultaneously, be explained in this framework?

We will also see there are operations for which the OLPA approach does not work.
Our main example is the operation $\frac{1}{2}L = \{ x \in \Sig^* : xy \in L, |x| = |y|\}$, which belongs to a general class of operations called proportional removals~\cite{Dom02}.
If  the OLPA approach fails for an operation, what does this imply about the nature and behaviour of this operation?
Are the state complexity problems for these operations ``harder'' to solve in general?
Studying operations for which the OLPA approach fails could be a fruitiful line of research. 

The purpose of this paper is to initiate a formal and general study of the OLPA approach. 
We define a class of regular operations, called ``uniform operations'', for which the OLPA approach provably works.
The class of uniform operations includes common operations such as reversal, star, power, concatenation, and all boolean operations (union, intersection, complement, etc.) but also includes more esoteric operations like cyclic shift~\cite{JiOk08} and shuffles on trajectories~\cite{DoSa04,MRS98}.
The class is also closed under composition, and thus includes combined operations like ``star-complement-star''~\cite{JiSh12}. 

We prove that for uniform operations, the worst-case state complexity can be determined by considering just a finite set of witnesses. For an $m$-ary operation, if the state complexity of the $j$-th input is at most $n_j$, then the worst-case state complexity of the output can be determined using $2(n_1+\dotsb+n_m)$ different witness languages, and by testing $2^mn_1\dotsb n_m$ different combinations of these witnesses.
Additionally, 
the same set of witnesses can be used for all uniform operations of a particular arity.

In the main sections of this paper, we will first state our definitions and prove our results in the special case of unary operations, before moving on to the general case.
While this is ultimately redundant, focusing on the unary case simplifies the notation and makes the definitions and results easier to digest.
The reader may find it useful to skip over the discussions of the general case on the first reading, and come back after they fully understand how the OLPA approach is formalized in the unary case.

To close the introduction, we give a history of the ideas behind the OLPA approach.
As mentioned, the key insight dates back to a 1978 paper of Sakoda and Sipser~\cite{SaSi78}.
They constructed languages wherein the alphabet letters were directed graphs representing behaviours of non-deterministic finite automata and two-way deterministic finite automata, with one letter for each possible behaviour. 
They used these languages to prove results on the complexity of conversions between different models of finite automata.

Perhaps the closest ancestor of our work is a 1990 paper of Ravikumar~\cite{Rav90}, who treated the ``Sakoda-Sipser technique'' as a ``systematic method to prove lower bounds on the size complexity of finite automata'', and applied it to five different problems, two of which were operational state complexity problems!
This is the first work we are aware of to present the OLPA approach as a general problem-solving method. 
Unfortunately, the field of state complexity was not very well-developed at the time, and Ravikumar seemingly did not realize the full generality and applicability of the approach in operational state complexity.
Furthermore, Ravikumar's use of the OLPA approach was less refined than the version we present; Ravikumar used $n$-state OLPA witnesses each with an alphabet of size $n^n$ as inputs to an $n$-ary operation, which is not guaranteed to maximize the state complexity of the operation. Our version of the approach would use an alphabet of size $n^n \dotsb n^n = (n^n)^n$.

In 1992, building off Ravikumar's work, Birget~\cite{Bir92} used this ``unrefined'' version of the OLPA approach to prove lower bounds on the state complexity of intersection and union.

In 1994, Yu, Zhuang and Salomaa~\cite{YZS94} published their seminal paper on the state complexities of basic operations.
Notably, even though Yu, Zhuang and Salomaa cited Ravikumar's work, they did \e{not} use or mention the ``Sakoda-Sipser technique'' anywhere in their paper, instead using various ad-hoc methods to prove lower bounds. 
The technique then seemingly faded into obscurity for a while.
It reappeared as the ``full automata technique'' in a 2006 paper of Yan~\cite{Yan06}, who credited Sakoda and Sipser for the idea, and applied it to nondeterministic finite automata and automata on infinite words ($\omega$-automata). Yan's paper is frequently cited in the field of $\omega$-automata, so the idea seems to have gained some currency there.

In the field of deterministic state complexity, the OLPA approach has made occasional past appearances.
Jir\'askov\'a and Okhotin~\cite{JiOk08} and later Domaratzki and Okhotin~\cite{DoOk09} used the OLPA approach to compute the exact worst-case complexity of the cyclic shift and power operations for small values. 
Brzozowski, Jir\'askov\'a, Liu, Rajasekaran, and Szyku\l{}a~\cite{BJLRS16} used the OLPA approach to obtain reachability results for the state complexity of shuffle.
Cho, Han, Ko and Salomaa~\cite{CHKS16} used an OLPA-like construction to establish lower bounds on the state complexity of some ``inversion'' operations. Interestingly, their construction includes unnecessary extra letters; perhaps these letters were added to somehow make the proof easier.

Outside the context of descriptional complexity, in 2002, Domaratzki, Kisman and Shallit~\cite{DKS02} used OLPA automata to enumerate the languages accepted by $n$-state automata.

In 2018, Caron, Hamel-De le court, Luque and Patrou~\cite{CHLP18} independently obtained many of the results in this paper using a different formalism. OLPA witnesses are called ``monsters'' in their work, and uniform operations are called ``depictable operations''. Their paper was submitted to arXiv just ten days after the first version of this paper was submitted.

\section{Preliminaries}
\label{sec:pre}
\subsection{Relations and Functions}
A \e{binary relation} $\rho$ between $X$ and $Y$ is a subset of $X \times Y$. If $\rho \subseteq X \times Y$ and $\tau \subseteq Y \times Z$, the \e{composition} of $\rho$ and $\tau$ is the relation 
\[ \rho\tau = \{(x,z) \in X \times Z : \text{ there exists } y \in Y \text{ such that } (x,y) \in \rho \text{ and } (y,z) \in \tau\}. \]
For $x \in X$ and $\rho \subseteq X \times Y$, the \e{image} of $x$ under $\rho$ is the set $x\rho = \{ y \in Y : (x,y) \in \rho\}$.
For $x \not\in X$ we define $x\rho = \emp$.
The \e{converse} of a binary relation $\rho \subseteq X \times Y$ is the relation $\rho^{-1} = \{ (y,x) : (x,y) \in \rho\} \subseteq Y \times X$.
The set $y\rho^{-1} = \{ x \in X : (x,y) \in \rho\}$ is called the \e{preimage} of $y$ under $\rho$. 

%
A \e{function} $f \co X \ra Y$ is a binary relation $f \subseteq X \times Y$ such that $|xf| = 1$ for all $x \in X$. Following our notation for binary relations, we write functions to the \e{right} of their arguments.
Composition of functions is defined by composing the corresponding relations. Thus the order of composition is \e{left-to-right}; in a composition $fg$, first $f$ is applied and then $g$.
A \e{transformation} of a set $X$ is a function $t \co X \ra X$, that is, a function from $X$ into itself.


\subsection{Languages and Automata}
\label{sec:def:aut}

A \e{finite automaton} (FA) is a tuple $\cA = (Q,\Sig,T,I,F)$ where $Q$ is a finite set of \e{states}, $\Sig$ is a finite set of \e{letters} called an \e{alphabet}, $T \subseteq Q \times \Sig \times Q$ is a set of \e{transitions}, $I \subseteq Q$ is a set of \e{initial states}, and $F \subseteq Q$ is a set of \e{final states}.
The triple $(Q,I,F)$ is called the \e{state configuration} of the automaton.

We now define a binary relation $T_w \subseteq Q \times Q$ for each $w \in \Sig^*$.
Define $T_\eps = \{(q,q) : q \in Q\}$; in terms of maps, this is the identity map on $Q$.
For $a \in \Sig$, define $T_a = \{ (p,q) \in Q \times Q : (p,a,q) \in T\}$.
For $w = a_1\dotsb a_k$ with $a_1,\dotsc,a_k \in \Sig$, define $T_w = T_{a_1} \dotsb T_{a_k}$.
The relation $T_w$ is called the \e{relation induced by $w$} or the \e{action of $w$}.
If $T_w$ is a transformation of the state set $Q$, it may also be called the \e{transformation induced by $w$}.
The set $\{T_w : w \in \Sig^*\}$ is a monoid under composition, called the \e{transition monoid} of $\cA$.



If $\cA = (Q,\Sig,T,I,F)$ is a finite automaton such that $|I| = 1$ and $T_a$ is a function for each $a \in \Sig$,  we say $\cA$ is \e{deterministic}. 
We abbreviate ``deterministic finite automaton'' to DFA. 
As a result of this definition, that all DFAs we consider in this paper are \e{complete} DFAs (that is, they have exactly one transition defined for each state-letter pair), and a finite automaton with an empty state set or with no initial state is not considered a DFA.

Let $\cA = (Q,\Sig,T,I,F)$ be an FA.
A word $w \in \Sig^*$ is \e{accepted} by $\cA$ if we have $IT_w \cap F \ne \emp$. If $\cA$ is a DFA with $I = \{i\}$, this condition becomes $iT_w \in F$. The \e{language} of $\cA$, denoted $\cL(\cA)$, is the set of all words it accepts. 
If $L$ is the language of $\cA$, we also say that \e{$\cA$ accepts $L$} and that \e{$\cA$ is an FA for $L$}.
Languages of FAs are called \e{regular languages}. 

A \e{regular operation} of arity $m$ is a function that takes $m$ regular languages as input and produces a regular language.
A \e{DFA operation} of arity $m$ is a function that takes $m$ DFAs as input and produces a DFA.
We say a regular operation $\Phi$ is \e{equivalent} to a DFA operation $\Psi$ if both operations have the same arity $m$, and for all $m$-tuples of DFAs $(\cD_1,\dotsc,\cD_m)$, we have $\cL((\cD_1,\dotsc,\cD_m)\Psi) = (\cL(\cD_1),\dotsc,\cL(\cD_m))\Phi$. 
In other words, they are equivalent if the DFA operation $\Psi$ is an ``implementation'' of the regular operation $\Phi$, in the sense that we can compute the output of $\Phi$ by taking arbitrary DFAs for the input languages, feeding them to $\Psi$, and taking the language of the output DFA.

In this paper we consider only DFA operations $\Psi$ that are \e{alphabet-preserving} in the following sense:
\bi
\item
An input $(\cD_1,\dotsc,\cD_m)$ is only valid if all the DFAs have the same alphabet.
\item
If $\Sig$ is the common alphabet of the input DFAs, then $\Sig$ will be the alphabet of the output DFA.
\ei
Furthermore, we consider only regular operations that are equivalent to an alphabet-preserving DFA operation.

\subsection{State Complexity}
\label{sec:def:stc}
A DFA for a regular language $L$ is \e{minimal} if it has the minimal number of states amongst all DFAs that accept $L$.
The \e{state complexity} of a regular language is the number of states in a minimal DFA accepting the language.
The state complexity of $L$ is denoted $\stc(L)$.

The notion of state complexity extends to regular operations.
Let $\Phi$ be a unary regular operation. The \e{state complexity of the operation $\Phi$} is the following function which takes a positive integer as input:
\[ n \mapsto \max \{\stc(L\Phi) : \stc(L) \le n\}. \]
That is, the state complexity of $\Phi$ is the worst-case state complexity of the output $L\Phi$, expressed as a function of the maximal allowed state complexity of the input $L$.
Note that $\max\{\stc(L\Phi) : \stc(L) \le n\}$ might not exist for all $n$; in such cases, the output of the function is $\infty$.

This idea generalizes to operations of higher arity. 
Let $\Phi$ be an $m$-ary regular operation. The state complexity of $\Phi$ is the following function which takes an $m$-tuple of positive integers as input:
\[ (n_1,\dotsc,n_m) \mapsto \max \{\stc((L_1,\dotsc,L_m)\Phi) : \stc(L_i) \le n_i, 1 \le i \le m\}. \]
The output is either a positive integer, or $\infty$ if the maximum does not exist. 

\subsection{Morphisms}
Let $\Sig$ and $\Gamma$ be alphabets.
A \e{morphism} is a function $\phi \co \Sig^* \ra \Gamma^*$ such that $(xy)\phi = (x\phi)(y\phi)$; in other words, a morphism is just a monoid homomorphism between two free monoids. To define a morphism $\phi \co \Sig^* \ra \Gamma^*$, it is sufficient to specify its values on letters from $\Sig$; the values on letters completely determine the values on words.

If $L \subseteq \Gamma^*$ is regular, then $L\phi^{-1}$ is regular.
To see this,
let $\phi \co \Sig^* \ra \Gamma^*$ be a morphism and let $\cB = (Q,\Gamma,T,i,F)$ be a DFA. We can construct a DFA for $\cL(\cB)\phi^{-1}$ as follows: let $\cB\phi^{-1} = (Q,\Sig,T',i,F)$, where $T' = \{(q,a,qT_{a\phi}) : q \in Q, a \in \Sig\}$.
Then it is easily verified that $\cL(\cB\phi^{-1}) = \cL(\cB)\phi^{-1}$.
We call $\cB\phi^{-1}$ the \e{inverse morphism DFA} of $\cB$ with respect to $\phi$.

Note that $\cB\phi^{-1}$ has the same state configuration as $\cB$. This will be useful for multiple reasons, but in particular it implies the following result for regular languages $L$ and $K$:
\bl
\label{lem:stc}
If $L = K\phi^{-1}$, then $\stc(L) \le \stc(K)$. 
\el

\section{Transformation Languages}
\label{sec:trans}
In this section, we formally define the witness languages that are used in the OLPA approach.

Fix a set $Q$ and let $\Sig$ be a set of transformations of $Q$.
For $i \in Q$ and $F \subseteq Q$, the \e{transformation language} $\Sig(i,F)$ is the language of the DFA $(Q,\Sig,T,i,F)$, where $T = \{ (q,t,qt) : q \in Q, t \in \Sig\}$. 
This DFA is called the \e{standard DFA} for the transformation language.

The set of all transformations of a set $Q$ is called the \e{full transformation monoid} on $Q$, and is denoted $\cT_Q$.
The \e{full transformation languages} of the form $\cT_Q(i,F)$ play in important role in the theory behind the OLPA approach.

Notice that the language $\cT_Q(i,F)$ has alphabet $\cT_Q$, and the standard DFA for $\cT_Q(i,F)$ has transitions $\{(q,t,qt) : q \in Q, t \in \cT_Q\}$.
This DFA has \e{one letter per transformation} of the state set $Q$, that is, one letter per possible action on the DFA's states.
Full transformation languages are the languages used as witnesses when applying the OLPA approach to unary operations.

Let $L$ be a regular language over $\Sig$ recognized by a DFA $\cD = (Q,\Sig,T,i,F)$. The \e{standard transformation morphism} of $L$ (with respect to $\cD$), denoted by $\phi_L \co \Sig^* \ra \cT_Q^*$, is defined by $a\phi_L = T_a$.
The following fact is easily verified:
\bp \label{prop:tfm} $L = \cT_Q(i,F)\phi_L^{-1}$. \ep


Full transformation languages do not suffice as OLPA witnesses for operations of arity greater than one.
When applying the OLPA approach to operations of arity $m$, we want to use an $m$-tuple $(\cD_1,\dotsc,\cD_m)$ of DFAs (where $\cD_j = (Q_j,\Sig,T_j,i_j,F_j)$ for $1 \le j \le m$) with the following property: for each $m$-tuple of transformations $(t_1 \co Q_1 \ra Q_1,\dotsc,t_m \co Q_m \ra Q_m)$, there exists a letter $a \in \Sig$ such that $a$ induces transformation $t_j$ in $\cD_j$ for $1 \le j \le m$. 
That is, we have one letter for every possible combination of actions across all the input DFAs.

For this purpose, we define \e{transformation tuple languages}.
Let $Q_1,\dotsc,Q_m$ be finite sets and let $\Sig$ be a subset of $\bT = \cT_{Q_1} \times \dotsb \times \cT_{Q_m}$.
For $j$ with $1 \le j \le m$, $i \in Q_j$, and $F \subseteq Q_j$, the \e{transformation tuple language} $\Sig_j(i,F)$ is the language of the DFA $(Q_j,\Sig,T,i,F)$ where $T = \{ (q,(t_1,\dotsc,t_m),qt_j) : q \in Q_j, (t_1,\dotsc,t_m) \in \Sig\}$.
This DFA is called the \e{standard DFA} of the transformation tuple language.
The \e{full transformation tuple languages} of the form $\bT_j(i,F)$ are used as OLPA witnesses in the case of $m$-ary operations.

There is a generalization of Proposition \ref{prop:tfm} for full transformation tuple languages.
Let $(L_1,\dotsc,L_m)$ be an $m$-tuple of regular languages over $\Sig$, where $L_j$ is recognized by the DFA $\cD_j = (Q_j,\Sig,T_j,i_j,F_j)$ for $1 \le i \le j$.
The \e{standard transformation tuple morphism} of $(L_1,\dotsc,L_m)$ (with respect to $(\cD_1,\dotsc,\cD_m)$), denoted by $\phi_{(L_1,\dotsc,L_m)} \co \Sig^* \ra (\cT_{Q_1} \times \dotsb \times \cT_{Q_m})^*$, is defined by $a\phi_{(L_1,\dotsc,L_m)} = ((T_1)_a,\dotsc,(T_m)_a)$.
As shorthand, let $\bT = \cT_{Q_1} \times \dotsb \times \cT_{Q_m}$ and let $\phi = \phi_{(L_1,\dotsc,L_m)}$.
\bp 
\label{prop:tftm} 
We have
$(L_1,\dotsc,L_m) = (\bT_1(i_1,F_1)\phi^{-1},\dotsc,\bT_m(i_m,F_m)\phi^{-1})$.
\ep
\bpf
It suffices to show for all $w \in \Sig^*$ that $w \in L_j \iff w\phi \in \bT_j(i_j,F_j)$.
Fix $j$ and let $(Q_j,\bT,T,i_j,F_j)$ be the standard DFA of $\bT_j(i_j,F_j)$.
Then we have
\begin{align*}
w \in L_j
\iff
i_j(T_j)_w \in F_j
\iff
i_jT_{w\phi} \in F_j
\iff
w\phi \in \bT_j(i_j,F_j),
\end{align*}
as required.

The second two-way implication may not be obvious. To see that it holds, first note that if $w$ is empty, then $(T_j)_w$ and $T_{w\phi}$ are both the identity map on $Q_j$. Otherwise, suppose $w = a_1\dotsb a_k$ with $a_1,\dotsc,a_k \in \Sig$.
We may write $w\phi = (a_1\phi) \dotsb (a_k\phi)$, and thus $T_{w\phi} = T_{a_1\phi} \dotsb T_{a_k\phi}$.
By definition, we have $a_i\phi = ((T_1)_{a_i},\dotsc,(T_m)_{a_i})$ for $1 \le i \le k$.
This $m$-tuple of transformations is a ``letter'' of the alphabet $\bT = \cT_{Q_1} \times \dotsb \cT_{Q_m}$.
Then $T_{a_i\phi}$ is the transformation of $Q_j$ induced by the ``letter'' $a_i\phi$.
By definition, this induced transformation is the map $q \mapsto q(T_j)_{a_i}$ for $q \in Q_j$.
Thus $T_{a_i\phi} = (T_j)_{a_i}$ for $1 \le i \le k$.
It follows that 
\[ T_{w\phi} = T_{a_1\phi} \dotsb T_{a_k\phi} = (T_j)_{a_1} \dotsb (T_j)_{a_k} = (T_j)_w. \]
Hence the implication holds. \qed
\epf

\section{Uniform Regular Operations}
\label{sec:uniform}
%

Our goal in this section is to define a large class of operations for which the OLPA approach works. 
The approach does not work for all regular operations; it is easy to come up with rather contrived examples of operations for which OLPA fails. 
Consider an operation which sends languages with one letter per action to the empty language, and acts as the identity on all other languages.
There are a few ways we could implement this operation as a DFA operation:
\bi
\item
If the input DFA has one letter per action, output a DFA with no final states. Otherwise, output the input DFA.
\item
If the input DFA has one letter per action, output a DFA in which the initial state is non-final and all the actions send the initial state to a sink state. Otherwise, output the input DFA.
\ei
The problem with this operation is that its behaviour is not ``uniform'' across all languages; it detects particular languages and has special behaviour for them.
In the first case, the operation does not behave uniformly on states: for most DFAs it preserves the final state set, but for DFAs with one letter per action it can change the final state set.
In the second case, the operation is not uniform on states \e{or} on actions: for most DFAs it preserves the state configuration and actions, but for DFAs with one letter per action, it can change whether the initial state is final, and also replace the actions by completely different actions.


We now attempt to formally define this idea of ``uniformity'' for unary operations.
Let $\Psi$ be a unary DFA operation.
We say $\Psi$ is \e{uniform} if for every pair of DFAs $\cA = (Q,\Sig,T_\cA,i,F)$ and $\cB = (Q,\Gamma,T_\cB,i,F)$ with the same state configuration, the image DFAs 
$\cA\Psi = (Q'_\cA,\Sig,T'_\cA,i'_\cA,F'_\cA)$ and 
$\cB\Psi = (Q'_\cB,\Gamma,T'_\cB,i'_\cB,F'_\cB)$ satisfy the following conditions:
\be
\item
$(Q'_\cA,i'_\cA,F'_\cA) = (Q'_\cB,i'_\cB,F'_\cB)$. 
\item
Whenever $(T_\cA)_a = (T_\cB)_b$ for $a \in \Sig$ and $b \in \Gamma$, we have
$(T'_\cA)_a = (T'_\cB)_b$.
\ee
We will say a unary regular operation $\Phi$ is \e{uniform} if there exists a uniform unary DFA operation equivalent to $\Phi$.

We can interpret this definition intuitively as follows.
The first condition says that the operation is uniform with respect to state configurations: if the operation is given two input DFAs with the same state configuration, it will produce two output DFAs with the same state configuration.
The second condition says that the operation is uniform with respect to actions: if the operation is given two input DFAs with the same state configuration and a common action, then it will produce two output DFAs with a common action, and furthermore the same letters which induce the common action in the input DFAs will induce the common action in the output DFAs.


The definition of uniformity is heavily dependent on DFAs.
Thus, it may come as a surprise that there is a simple and purely language-theoretic characterization of uniformity. 
A morphism $\phi \co \Sig^* \ra \Gamma^*$ is \e{1-uniform} if it maps letters to letters.
\bp
\label{prop:uniform}
Let $L$ and $K$ be regular languages over $\Sig$ and $\Gamma$ respectively.
The following are equivalent:
\be
\item
\label{prop:uniform:op}
The regular operation $\Phi$ is uniform.
\item
\label{prop:uniform:morphism}
For all $1$-uniform morphisms $\phi \co \Sig^* \ra \Gamma^*$, if $L = K\phi^{-1}$ then $L\Phi = K\Phi\phi^{-1}$.
\ee
\ep
\bpf
$(1) \implies (2)$: 
Since $\Phi$ is uniform, there is a uniform DFA operation $\Psi$ equivalent to $\Phi$.
Fix a 1-uniform morphism $\phi \co \Sig^* \ra \Gamma^*$ such that $L = K\phi^{-1}$.
Let $\cA = (Q_\cA,\Sig,T_\cA,i_\cA,F_\cA)$ be a DFA for $L$, and let $\cB = (Q_\cB,\Gamma,T_\cB,i_\cB,F_\cB)$ be a DFA for $K$.
We write $w_\cA$ for $(T_{\cA})_w$, and $w_\cB$ for $(T_\cB)_w$.

Note that we can choose our DFAs so that they have the same state configuration.
This follows from the fact that $L = K\phi^{-1}$, and thus we can take $\cA = \cB\phi^{-1}$ which has the same state configuration as $\cB$.
Henceforth write $Q_\cA = Q_\cB = Q$, $i_\cA = i_\cB = i$, and $F_\cA = F_\cB = F$.

Let $\cA\Psi = \cA' = (Q'_\cA,\Sig,T'_\cA,i',F'_\cA)$ and let $\cB\Psi = \cB' = (Q'_\cB,\Gamma,T'_\cB,i'_\cB,F'_\cB)$.
Write $w'_\cA$ for $(T'_\cA)_w$ and $w'_\cB$ for $(T'_\cB)_w$.
The DFA $\cA'$ recognizes $L\Phi$, and the DFA $\cB'$ recognizes $K\Phi$.
By the uniformity of $\Psi$, we can write $Q'_\cA = Q'_\cB = Q'$, $i'_\cA = i'_\cB = i'$, and $F'_\cA = F'_\cB = F'$.

Now, we want to show that $L\Phi = K\Phi\phi^{-1}$.
Since $\cA = \cB\phi^{-1}$, for all $q \in Q$ and $a \in \Sig$ we have $qa_\cA = q(a\phi)_\cB$ by definition.
Thus $a_\cA$ and $(a\phi)_\cB$ are equal as transformations of $Q$ for all $a \in \Sig$.
By the uniformity of $\Psi$,
$a'_\cA$ and $(a\phi)'_\cB$ are equal as transformations of $Q'$.
It follows that 
$w'_\cA$ and $(w\phi)'_\cB$ are equal as transformations of $Q'$ for all $w \in \Sig^*$.
Hence we have
{\small
\[ w \in L\Phi \iff i'w'_\cA \in F' \iff i'(w\phi)'_\cB \in F' \iff w\phi \in K\Phi \iff w \in K\Phi\phi^{-1}. \]
}
This proves that $L\Phi = K\Phi\phi^{-1}$.

$(2) \implies (1)$:
We are given a regular operation $\Phi$.
We want to produce a uniform DFA operation $\Psi$ such that for all DFAs $\cA$, we have $\cL(\cA\Psi) = \cL(\cA)\Phi$.

Fix an $n$-state DFA $\cA = (Q,\Sig,T,i,F)$ and let $L$ be its language. 
We define $\cA\Psi$ as follows.
By Proposition \ref{prop:tfm}, we have $L = \cT_Q(i,F)\phi_L^{-1}$, where $\phi_L \co \Sig^* \ra \cT_Q^*$ is the standard transformation morphism of $L$.
By assumption, we then have $L\Phi = \cT_Q(i,F)\Phi\phi_L^{-1}$.
Let $\cD' = (Q',\cT_Q,T',i',F')$ be a minimal DFA for $\cT_Q(i,F)\Phi$,
and set $\cA\Psi = \cD'\phi^{-1}_L$.

It is clear that we have $\cL(\cA\Psi) = \cT_Q(i,F)\Phi\phi^{-1}_{\cL(\cA)} =  \cL(\cA)\Phi$ as required.
To see that $\Psi$ is uniform, fix DFAs $\cA = (Q,\Sig,T_\cA,i,F)$ and $\cB = (Q,\Gamma,T_\cB,i,F)$.
We compute the images 
$\cA\Psi = \cA' = (Q'_\cA,\Sig,T'_\cA,i'_\cA,F'_\cA)$
and
$\cB\Psi = \cB' = (Q'_\cB,\Gamma,T'_\cB,i'_\cB,F'_\cB)$.
Now, let $\cD' = (Q',\cT_Q,T'_\cD,i',F')$ be the minimal DFA for $T_Q(i,F)\Phi$.
By definition, we have 
$\cA' = \cD'\phi^{-1}_{\cL(\cA)}$
and
$\cB' = \cD'\phi^{-1}_{\cL(\cB)}$.
So $\cA'$ and $\cB'$ both have the same state configuration as $\cD'$.
It follows that $(Q'_\cA,i'_\cA,F'_\cA) =(Q'_\cB,i'_\cB,F'_\cB)$, as required.

Next, fix $a \in \Sig$ and $b \in \Gamma$ such that $(T_\cA)_a = (T_\cB)_b$.
We have
$q(T'_\cA)_a = q(T'_\cD)_{a\phi_{\cL(\cA)}}$ for all $q$.
Also,
$q(T'_\cB)_b = q(T'_\cD)_{b\phi_{\cL(\cB)}}$ for all $q$.
By the definition of the standard transformation morphism, we have $a\phi_{\cL(\cA)} = b\phi_{\cL(\cB)}$, since $(T_\cA)_a = (T_\cB)_b$.
It follows that $(T'_\cA)_a = (T'_\cB)_b$, as required.
Thus $\Psi$ is uniform.
\qed
\epf

Now that we have established the definition of uniformity and the language-theoretic characterization for unary regular operations, we turn to operations of higher arity. 
Let $\Psi$ be an $m$-ary DFA operation.
We say $\Psi$ is \e{uniform} if for every pair of $m$-tuples of DFAs $(\cA_1,\dotsb,\cA_m)$ and $(\cB_1,\dotsc,\cB_m)$,
where for each $j$ with $1 \le j \le m$, the DFAs $\cA_j$ and $\cB_j$ have the same state configuration, the DFA $\cA_j$ has alphabet $\Sig$ and transition set $T_{\cA_j}$, and the DFA $\cB_j$ has transition set $T_{\cB_j}$ and alphabet $\Gamma$; the image DFAs $\cA = (\cA_1,\dotsb,\cA_m)\Psi$ and $\cB = (\cB_1,\dotsc,\cB_m)\Psi$ have transition sets $T_{\cA}$ and $T_{\cB}$ respectively, and the following conditions hold:
\be
\item
The image DFAs $\cA$ and $\cB$ have the same state configuration.
\item
If there exist letters $a \in \Sig$ and $b \in \Gamma$ such that 
$(T_{\cA_j})_a = (T_{\cB_j})_b$ for each $j$ with $1 \le j \le m$, then 
$(T_\cA)_a = (T_\cB)_b$.
\ee
There is a corresponding language-theoretic characterization of the general definition of uniformity.
\bp
\label{prop:uniform-general}
Let $(L_1,\dotsc,L_m)$ and $(K_1,\dotsc,K_m)$ be $m$-tuples of regular languages, where $L_j$ is a language over $\Sig$ and $K_j$ is a language over $\Gamma$ for $1 \le j \le m$.
The following are equivalent:
\be
\item
\label{prop:uniform-general:op}
The $m$-ary regular operation $\Phi$ is uniform.
\item
\label{prop:uniform-general:morphism}
For all $1$-uniform morphisms $\phi \co \Sig^* \ra \Gamma^*$, if $L_j = K_j\phi^{-1}$ for $1 \le j \le m$, then $(L_1,\dotsc,L_m)\Phi = (K_1,\dotsc,K_m)\Phi\phi^{-1}$.
\ee
\ep
The proof is very similar to the proof of Proposition \ref{prop:uniform}, except the general definition of uniformity is used and full transformation tuple languages are used instead of full transformation languages. 
\bpf
$(1) \implies (2)$: 
Since $\Phi$ is uniform, there is a uniform DFA operation $\Psi$ equivalent to $\Phi$.
Fix a 1-uniform morphism $\phi \co \Sig^* \ra \Gamma^*$ such that $L_j = K_j\phi^{-1}$ for $1 \le j \le m$.
We want to show that $(L_1,\dotsc,L_m)\Phi = (K_1,\dotsc,K_m)\Phi\phi^{-1}$.

Since $L_j = K_j\phi^{-1}$, for each $j$ we can find a DFA $\cA_j$ for $L_j$ and a DFA $\cB_j$ for $K_j$ such that $\cA_j = \cB_j\phi^{-1}$.
Each pair of DFAs $\cA_j$ and $\cB_j$ has a common state configuration $(Q_j,i_j,F_j)$.
For $1 \le j \le m$, let $\cA_j = (Q_j,\Sig,T_{\cA_j},i_j,F_j)$ be the DFA for $L_j$ and let $\cB_j = (Q_j,\Gamma,T_{\cB_j},i_j,F_j)$ be the DFA for $K_j$.
By the uniformity of $\Psi$, the image DFAs $\cA = (\cA_1,\dotsc,\cA_m)\Psi$ and $\cB = (\cB_1,\dotsc,\cB_m)\Psi$ have a common state configuration $(Q,i,F)$. Write $\cA = (Q,\Sig,T_\cA,i,F)$ and $\cB = (Q,\Sig,T_\cB,i,F)$.

Since $\cA_j = \cB_j\phi^{-1}$ for $1 \le j \le m$, for all $q \in Q_j$ and $a \in \Sig$, we have $q(T_{\cA_j})_a = q(T_{\cB_j})_{a\phi}$ by definition. Thus $(T_{\cA_j})_a = (T_{\cB_j})_{a\phi}$ for all $a \in \Sig$ and all $1 \le j \le m$. By the uniformity of $\Psi$, it follows that $(T_\cA)_a = (T_\cB)_{a\phi}$.
Hence $(T_\cA)_w = (T_\cB)_{w\phi}$ for all $w \in \Sig^*$.
Thus we have
\begin{align*}
w \in (L_1,\dotsc,L_m)\Phi &\iff i(T_\cA)_w \in F \iff i(T_\cB)_{w\phi} \in F \\
&\iff w\phi \in (K_1,\dotsc,K_m)\Phi \iff w \in (K_1,\dotsc,K_m)\Phi\phi^{-1}.
\end{align*}
This proves that $(L_1,\dotsc,L_m)\Phi = (K_1,\dotsc,K_m)\Phi\phi^{-1}$.

$(2) \implies (1)$: 
We want to produce a uniform $m$-ary DFA operation $\Psi$ such that for all tuples of DFAs $(\cA_1,\dotsc,\cA_m)$ over a common alphabet, we have 
$\cL((\cA_1,\dotsc,\cA_m)\Psi) = (\cL(\cA_1),\dotsc,\cL(\cA_m))\Phi$.

Fix a tuple $(\cA_1,\dotsc,\cA_m)$ of DFAs over $\Sig$, where $\cA_j$ has state configuration $(Q_j,i_j,F_j)$, and let $L_j = \cL(\cA_j)$ for $1 \le j \le m$. 
We define the image $\cA = (\cA_1,\dotsc,\cA_m)\Psi$ as follows.
By Proposition \ref{prop:tftm} we have $L_j = \bT_j(i_j,F_j)\phi^{-1}_{(L_1,\dotsc,L_m)}$, where $\phi_{(L_1,\dotsc,L_m)}$ is the standard transformation tuple morphism of $(L_1,\dotsc,L_m)$ with respect to $(\cA_1,\dotsc,\cA_m)$, and $\bT = \cT_{Q_1} \times \dotsc \times \cT_{Q_m}$.
Let $\cD$ be a minimal DFA for $(\bT_1(i_1,F_1),\dotsc,\bT_m(i_m,F_m))\Phi$, and set $(\cA_1,\dotsc,\cA_m)\Psi = \cD\phi^{-1}_{(L_1,\dotsc,L_m)}$.

We claim that $\cL((\cA_1,\dotsc,\cA_m)\Psi) = (\cL(\cA_1),\dotsc,\cL(\cA_m))\Phi$.
Indeed, since
$L_j = \bT_j(i_j,F_j)\phi^{-1}_{(L_1,\dotsc,L_m)}$, 
we have
\[ (L_1,\dotsc,L_m)\Phi = 
(\bT_1(i_1,F_1),\dotsc,\bT_m(i_m,F_m))\Phi\phi^{-1}_{(L_1,\dotsc,L_m)}. \]
It follows that
{\small
\begin{align*}
\cL((\cA_1,\dotsc,\cA_m)\Psi
&= \cL(\cD\phi^{-1}_{(L_1,\dotsc,L_m)})
= (\bT_1(i_1,F_1),\dotsc,\bT_m(i_m,F_m))\Phi\phi^{-1}_{(L_1,\dotsc,L_m)}\\
&= (L_1,\dotsc,L_m)\Phi
= (\cL(\cA_1),\dotsc,\cL(\cA_m))\Phi,
\end{align*}
}
as required.

To see that $\Psi$ is uniform, fix $m$-tuples of DFAs $(\cA_1,\dotsc,\cA_m)$ and $(\cB_1,\dotsc,\cB_m)$ such that for $1 \le j \le m$, the DFAs $\cA_j$ and $\cB_j$ have the same state configuration $(Q_j,i_j,F_j)$, the DFA $\cA_j$ has alphabet $\Sig$ and transition set $T_{\cA_j}$, and the DFA $\cB_j$ has alphabet $\Gamma$ and transition set $T_{\cB_j}$.
Write $(\cA_1,\dotsc,\cA_m)\Psi = \cA = (Q_\cA,\Sig  ,T_\cA,i_\cA,F_\cA)$
and   $(\cB_1,\dotsc,\cB_m)\Psi = \cB = (Q_\cB,\Gamma,T_\cB,i_\cB,F_\cB)$.
Let $\cD$ be a minimal DFA for $(\bT_1(i_1,F_1),\dotsc,\bT_m(i_m,F_m))\Phi$ used in the definition of $\Psi$. 
Then $\cA$ and $\cB$ are both inverse morphism DFAs constructed from $\cD$, so they both have the same state configuration as $\cD$.
Write $(Q,i,F)$ for this common state configuration.

It remains to show that whenever we have $a \in \Sig$ and $b \in \Gamma$ such that $(T_{\cA_j})_a = (T_{\cB_j})_b$ for $1 \le j \le m$, it follows that $(T_\cA)_a = (T_\cB)_b$.
Fix $a \in \Sig$ and $b \in \Gamma$ with this property.
Write $\phi_\cA$ as shorthand for $\phi_{(\cL(\cA_1),\dotsc,\cL(\cA_m)}$, and write $\phi_\cB$ for $\phi_{(\cL(\cB_1),\dotsc,\cL(\cB_m)}$. 
By definition, we have $\cA = \cD\phi^{-1}_\cA$ and $\cB = \cD\phi^{-1}_\cB$.
Let $T_\cD$ be the transition set of $\cD$.
Then for $q \in Q$, we have $q(T_\cA)_a = q(T_\cD)_{a\phi_\cA}$ and $q(T_\cB)_b = q(T_{\cD})_{b\phi_\cB}$.
By the definition of the standard transformation tuple morphism, we have
$a\phi_\cA = ((T_{\cA_1})_a,\dotsc,(T_{\cA_m})_a)$ and $b\phi_\cB = ((T_{\cB_1})_b,\dotsc,(T_{\cB_m})_b)$.
But we are assuming that $(T_{\cA_j})_a = (T_{\cB_j})_b$ for $1 \le j \le m$, so in fact $a\phi_\cA = b\phi_\cB$.
It then follows that $(T_\cA)_a = (T_\cB)_b$, as required.

This proves that $\Psi$ is uniform, and thus $\Phi$ is uniform, since it is equivalent to a uniform DFA operation. \qed
\epf

\section{The Main Theorem}
\label{sec:main}
The goal of this section is to prove that the OLPA approach works for all uniform operations.
Thanks to Proposition \ref{prop:uniform}
and its generalization in Proposition \ref{prop:uniform-general}, 
this is not especially difficult. 

First we consider unary operations.
The following lemma formalizes a ``weak'' version of the OLPA approach for unary operations. The short proof contains all the essential ideas, but the expression it gives for the state complexity function is not practical, since it involves taking a maximum over all sets of size $n$. Obtaining a practical expression for the state complexity function is just a technical matter that we will deal with after this proof.
\bl
\label{lem:olpa}
Let $\Phi$ be a uniform unary regular operation. Let $L$ be a regular language recognized by a DFA $(Q,\Sig,T,i,F)$. Then $\stc(L\Phi) \le \stc(\cT_Q(i,F)\Phi)$. In particular, the state complexity of $\Phi$ is given by the following function:
\[ n \mapsto \max\{ \stc(\cT_Q(i,F)\Phi) : |Q| = n, i \in Q, F \subseteq Q\}. \]
\el
\bpf
Fix  $L$, and recall that $L = \cT_Q(i,F)\phi_L^{-1}$, where $\phi_L$ is the standard transformation morphism of $L$.
Since $\Phi$ is uniform, we have $L\Phi = \cT_Q(i,F)\Phi\phi_L^{-1}$ by Proposition \ref{prop:uniform}.
By Lemma \ref{lem:stc}, we have $\stc(L\Phi) \le \stc(\cT_Q(i,F)\Phi)$ as required. \qed
\epf

Now, we show that to compute the state complexity function, it suffices to just consider the set $\{1,\dotsc,n\}$ instead of all sets of size $n$. Furthermore, we may assume that $i = 1$, and that $F$ is either $\{1,\dotsc,k\}$ or $\{n-k+1,\dotsc,n\}$ for some $k \le n$. Thus it suffices to just check $2n$ OLPA witnesses.
Write $\cT_n$ for $\cT_{{\{1,\dotsc,n\}}}$, let $F_{n,k,1} = \{1,\dotsc,k\}$, and let $F_{n,k,0} = \{n-k+1,\dotsc,n\}$.
\newpage
\bt
\label{thm:olpa}
Let $\Phi$ be a uniform unary regular operation. Let $L$ be a regular language recognized by a DFA $(Q,\Sig,T,i,F)$.
Then $\stc(L\Phi) \le \stc(\cT_n(1,F_{n,k,j})\Phi)$, where $n=|Q|$, $k=|F|$, and $j$ is defined to be $1$ if $i \in F$ and $0$ if $i \not\in F$.
The state complexity of $\Phi$ is the following function:
\[ n \mapsto \max \{ \stc(\cT_{n}(1,F_{n,k,j})\Phi) : 0 \le j \le 1, 0+j \le k \le n-1+j\}. \]
\et
\bpf
We know from Lemma \ref{lem:olpa} that $\stc(L\Phi) \le \stc(\cT_Q(i,F)\Phi)$.
Let us prove that $\stc(\cT_Q(i,F)\Phi) \le \stc(\cT_n(1,F_{n,k,j})\Phi)$, with $n$, $k$ and $j$ defined as in the statement of the theorem. 

By Lemma \ref{lem:stc} and the uniformity of $\Phi$, it suffices to exhibit a morphism $\phi \co \cT_Q^* \ra \cT_n^*$ such that 
$\cT_Q(i,F) = \cT_n(1,F_{n,k,j})\phi^{-1}$.
To define $\phi$, first we define a bijection $\beta \co Q \ra \{1,\dotsc,n\}$.
We take $\beta$ to be a bijection with the following properties: $i\beta = 1$ and $F\beta = F_{n,k,j}$. 
The remaining elements of $Q$ are mapped to the remaining elements of $\{1,\dotsc,n\}$ arbitrarily.
Note that this definition of $\beta$ is only possible because of our choice of the parameter $j$. Indeed, we are mapping $i\beta$ to $1$, so if $i \in F$ then we better have $1 \in F_{n,k,j}$; but $i \in F$ implies $j = 1$ and thus $F_{n,k,j} = \{1,\dotsc,k\}$. On the other hand, if $i \not\in F$ then we better have $1 \not\in F_{n,k,j}$, but in this case we have $j = 0$ giving $F_{n,k,j} = \{n-k+1,\dotsc,n\}$.
Given this definition of $\beta$, for each $t \co Q \ra Q$, we define $t\phi$ be the transformation of $\{1,\dotsc,n\}$ that sends $m$ to $m\beta^{-1}t\beta$.

We show that $w \in \cT_Q(i,F)$ if and only if $w \in \cT_n(1,F_{n,k,j})\phi^{-1}$.
Let $w = t_1\dotsb t_k$ for $t_1,\dotsc,t_k \in \cT_Q$.
\begin{align*}
w \in \cT_Q(i,F) 
&\iff
iw \in F
\iff
1\beta^{-1} w \in F
\iff
1\beta^{-1} w \beta \in F\beta\\
&\iff
1 \beta^{-1} w \beta \in F_{n,k,j}
\iff
1 \beta^{-1} t_1t_2 \dotsb t_k \beta \in F_{n,k,j}\\
&\iff
1 \beta^{-1} t_1 \beta \beta^{-1} t_2 \beta \dotsb \beta^{-1} t_k \beta \in F_{n,k,j}\\
&\iff
1 (t_1\phi)(t_2\phi)\dotsb(t_k\phi) \in F_{n,k,j}
\iff
1(w\phi) \in F_{n,k,j}\\
&\iff
w\phi \in \cT_n(1,F_{n,k,j})
\iff
w \in \cT_n(1,F_{n,k,j})\phi^{-1}. 
\end{align*}
Thus $\cT_Q(i,F) = \cT_n(1,F_{n,k,j})\phi^{-1}$, as required. This completes the proof. \qed
\epf



We now consider uniform operations of arbitrary arity. The proof strategies in this case are much the same, except full transformation tuple languages are used as witnesses, rather than full transformation languages.
\bl
\label{lem:olpa-general}
Let $\Phi$ be a uniform $m$-ary regular operation. 
Let $(L_1,\dotsc,L_m)$ be regular languages, where $L_j$ is recognized by a DFA $(Q_j,\Sig,T_j,i_j,F_j)$.
Let $\bT = \cT_{Q_1} \times \dotsb \times \cT_{Q_m}$.
Then $\stc((L_1,\dotsc,L_m)\Phi) \le \stc((\bT_1(i_1,F_1),\dotsc,\bT_m(i_m,F_m))\Phi)$.
\el
\bpf
Fix $(L_1,\dotsc,L_m)$, and recall from Proposition \ref{prop:tftm} that  
$(L_1,\dotsc,L_m) = (\bT_1(i_1,F_1)\phi^{-1},\dotsc,\bT_m(i_m,F_m)\phi^{-1})$, where the morphism $\phi = \phi_{(L_1,\dotsc,L_m)}$ is the standard transformation tuple morphism of $(L_1,\dotsc,L_m)$.
Since $\Phi$ is uniform, we have 
\[ (L_1,\dotsc,L_m)\Phi = (\bT_1(i_1,F_1),\dotsc,\bT_m(i_m,F_m))\Phi\phi^{-1},\] 
by Proposition \ref{prop:uniform-general}.
Then by Lemma \ref{lem:stc}, we have 
\[ \stc((L_1,\dotsc,L_m)\Phi) \le \stc((\bT_1(i_1,F_1),\dotsc,\bT_m(i_m,F_m)\Phi),\] 
as required. \qed
\epf
As before, it suffices to only check a finite number of witnesses.
Recall that we defined $\cT_n = \cT_{{\{1,\dotsc,n\}}}$, and for $k \le n$ we defined $F_{n,k,1} = \{1,\dotsc,k\}$ and $F_{n,k.0} = \{n-k+1,\dotsc,n\}$.

\bt[The ``Fundamental Theorem of the OLPA Approach'']
\label{thm:olpa-general}
Let $\Phi$ be a uniform $m$-ary regular operation. 
Let $(L_1,\dotsc,L_m)$ be regular languages, where $L_j$ is recognized by a DFA $(Q_j,\Sig,T_j,i_j,F_j)$.
Let $n_j = |Q_j|$ and let $\bT = \cT_{n_1} \times \dotsc \cT_{n_m}$.
Then we have 
\[ \stc((L_1,\dotsc,L_m)\Phi) \le \stc((\bT_1(1,F_{n_1,k_1,\ell_1}),\dotsc,\bT_m(1,F_{n_m,k_m,\ell_m}))\Phi), \]
where $k_j=|F_j|$, and $\ell_j$ is defined to be $1$ if $i_j \in F_j$ and $0$ if $i_j \not\in F_j$.
The state complexity of $\Phi$ is the function 
\[ (n_1,\dotsc,n_m) \mapsto \max \stc((\bT_1(1,F_{n_1,k_1,\ell_1}),\dotsc,\bT_m(1,F_{n_m,k_m,\ell_m}))\Phi), \]
where the maximum is taken over all possible values in the following ranges:
$1 \le j \le m$,
$0 \le \ell_j \le 1$,
and 
$0+\ell_j \le k_j \le n_j-1+\ell_j$.
\et
To compute the worst-case state complexity of an $m$-ary operation for $m$ input DFAs of sizes $n_1$ through $n_m$, we use $2(n_1 + \dotsb + n_m)$ different languages, each with an alphabet of size $n_1^{n_1}\dotsb n_m^{n_m}$. The number of input $m$-tuples that must be tested is $2^mn_1\dotsb n_m$, since for the $j$-th component there are $2n_j$ choices.
\bpf
Define $\bT' = \cT_{Q_1} \times \dotsc \times \cT_{Q_m}$.
We know from Lemma \ref{lem:olpa-general} that 
\[ \stc((L_1,\dotsc,L_m)\Phi) \le \stc((\bT'_1(i_1,F_1),\dotsc,\bT'_m(i_m,F_m))\Phi). \]
Let us prove the following:
{\small
\[ \stc((\bT'_1(i_1,F_1),\dotsc,\bT'_m(i_m,F_m))\Phi) \le 
\stc((\bT_1(1,F_{n_1,k_1,\ell_1}),\dotsc,\bT_m(1,F_{n_m,k_m,\ell_m}))\Phi), 
\]
}
where $n_j,k_j,\ell_j$ for $1 \le j \le m$ are defined as in the statement of the theorem.

By Lemma \ref{lem:stc} and the uniformity of $\Phi$, it suffices to exhibit a morphism 
$\phi \co (\bT')^* \ra \bT^*$
such that 
$\bT'_j(i_j,F_j) = \bT_j(1,F_{n_j,k_j,\ell_j})\phi^{-1}$
for $1 \le j \le m$.
To define $\phi$, first we define bijections $\beta_j \co Q_j \ra \{1,\dotsc,n_j\}$ for $1 \le j \le m$.
As in the proof of Theorem \ref{thm:olpa},
we take each $\beta_j$ to be a bijection with the following properties: $i_j\beta = 1$ and $F_j\beta = F_{n_j,k_j,\ell_j}$. 
The remaining elements of $Q_j$ are mapped to the remaining elements of $\{1,\dotsc,n_j\}$ arbitrarily.
Then for each tuple $(t_1,\dotsc,t_m) \in \bT'$, we define $(t_1,\dotsc,t_m)\phi$ to be the transformation tuple $(\beta_1^{-1}t_1\beta_1,\dotsc,\beta_m^{-1}t_m\beta_m)$, which lies in $\bT$.

Now, for $1 \le j \le m$, we show that
$w \in \bT'_j(i_j,F_j) \iff w\phi \in \bT_j(1,F_{n_j,k_j,\ell_j})$.
Let $w = (t_{1,1},\dotsc,t_{m,1})\dotsb(t_{1,k},\dotsc,t_{m,k})$, where each of these transformation tuples lies in $\bT'$.
\begin{align*}
w \in \bT'_j(i_j,F_j)
&\iff
i_jw \in F_j
\iff
i_jt_{j,1} t_{j,2} \dotsb t_{j,k} \in F_j\\
&\iff 
1 \beta_j^{-1} t_{j,1} t_{j,2} \dotsb t_{j,k} \beta_j \in F_{n_j,k_j,\ell_j}\\
&\iff 
1 \beta_j^{-1} t_{j,1} \beta_j \beta_j^{-1} t_{j,2} \beta_j \dotsb \beta_j^{-1} t_{j,k} \beta_j \in F_{n_j,k_j,\ell_j}\\
&\iff 
1 w\phi \in F_{n_j,k_j,\ell_j}
\iff
w\phi \in \bT_j(1,F_{n_j,k_j,\ell_j}).
\end{align*}
Thus $\bT'_j(i_j,F_j) = \bT_j(1,F_{n_j,k_j,\ell_j})\phi^{-1}$, as required. \qed
\epf


\section{Examples of Uniform and Non-Uniform Operations}
\label{sec:examples}
In this section, we prove that a number of common operations (as well as a few more esoteric ones) are uniform, and that the class of uniform operations is closed under composition. We also give some examples of non-uniform operations. 

\subsection{Uniform Operations}

First we consider a class of operations called \e{shuffles on trajectories}~\cite{DoSa04,MRS98}. 
Operations in this class include shuffle, literal shuffle, balanced literal shuffle, insertion, balanced insertion, concatenation, and anti-concatenation~\cite[Remark 3.1]{MRS98}. 
We denote the shuffle of languages $L$ and $L'$ along the set of trajectories $X \subseteq \{0,1\}^*$ by $L \shu_X L'$.
The shuffle on trajectories $L \shu_X L'$ is regular if and only if $X$ is regular~\cite[Theorem 5.1]{MRS98}.
For the definition of $L \shu_X L'$, see~\cite[Section 3]{MRS98}; for the following proof we only need to know the DFA construction.
\bp
For all regular languages $X \subseteq \{0,1\}^*$,
the shuffle on trajectories operation $(L,L') \mapsto L \shu_X L'$ is uniform.
\ep
\bpf
Following~\cite{MRS98}, we define a DFA operation $\Psi$ equivalent to the shuffle on trajectories operation.
Let $\cD_1 = (Q_1,\Sig,T_1,i_1,F_1)$ and $\cD_2 = (Q_2,\Sig,T_2,i_2,F_2)$ be arbitrary DFAs.
Let $\cD_X = (Q_X,\{0,1\},T_X,i_X,F_X)$ be a DFA for the set of trajectories $X$.
We set $(\cD_1,\cD_2)\Psi$ to be the determinization of the following FA $\cD$.
The FA $\cD$ has state set $Q_1 \times Q_X \times Q_2$, alphabet $\Sig$, initial state $(i_1,i_X,i_2)$, final state set $F_1 \times F_X \times F_2$, and transition set $T$ defined as follows: for each $a \in \Sig$, we have
\[ (q_1,q_X,q_2)T_a = \{ (q_1(T_1)_a,q_X(T_X)_0,q_2),(q_1,q_X(T_X)_1,q_2(T_2)_a)\}. \]
It was proved in~\cite[Theorem 5.1]{MRS98} that $\cL((\cD_1,\cD_2)\Psi) = \cL(\cD_1) \shu_X \cL(\cD_2)$.

Let $(\cA_1,\cA_2)$ and $(\cB_1,\cB_2)$ be pairs of DFAs such that for $1 \le j \le 2$, the DFAs $\cA_j$ and $\cB_j$ have the same state configuration $(Q_j,i_j,F_j)$, the DFA $\cA_j$ has alphabet $\Sig$ and transition set $T_{\cA_j}$, and the DFA $\cB_j$ has alphabet $\Gamma$ and transition set $T_{\cB_j}$.

It is clear that the image DFAs $\cA = (\cA_1,\cA_2)\Psi$ and $\cB = (\cB_1,\cB_2)\Psi$ will have the same state configuration.
Additionally, by inspecting the definitions of the transition sets $T_\cA$ of $\cA$ and $T_\cB$ of $\cB$, it is clear that if $(T_{\cA_j})_a = (T_{\cB_j})_b$ for $a \in \Sig$, $b \in \Gamma$ and $1 \le j \le 2$, then $(T_\cA)_a = (T_\cB)_b$.
Indeed, let $S \subseteq Q_1 \times Q_X \times Q_2$. Then we have
\begin{align*}
S(T_\cA)_a
&= 
\bigcup_{(q_1,q_X,q_2) \in S} \{(q_1,q_X,q_2)\}(T_{\cA})_a\\
&=
\bigcup_{(q_1,q_X,q_2) \in S} 
\{ (q_1(T_{\cA_1})_a,q_X(T_X)_0,q_2),(q_1,q_X(T_X)_1,q_2(T_{\cA_2})_a)\}\\
&=
\bigcup_{(q_1,q_X,q_2) \in S} 
\{ (q_1(T_{\cB_1})_b,q_X(T_X)_0,q_2),(q_1,q_X(T_X)_1,q_2(T_{\cB_2})_b)\}\\
&= 
\bigcup_{(q_1,q_X,q_2) \in S} \{(q_1,q_X,q_2)\}(T_{\cB})_b
=
S(T_\cB)_b.
\end{align*}
Thus $\Psi$ is uniform, and it follows that the shuffle on trajectories operation is uniform. \qed
\epf
The above proof illustrates the fact that once one understands the definition of uniformity, it is often easy to determine whether an operation is uniform just by inspecting the DFA construction.
There are no difficult ideas in this proof; it is just a statement of a DFA construction and a rudimentary calculation.

One can also use the language-theoretic characterization of uniformity to prove that operations are uniform. Typically, proofs using the language-theoretic characterization require somewhat more thought to write and read, but are shorter and have less of the ``boilerplate'' needed for DFA-based proofs. The rest of our uniformity proofs will use the language-theoretic characterization.
\bp
The reversal operation $L \mapsto L^R$ is uniform.
\ep
\bpf
Fix a 1-uniform morphism $\phi \co \Sig^* \ra \Gamma^*$ and suppose $L = K\phi^{-1}$. 
Since $\phi$ is 1-uniform, we have $(w^R)\phi = (w\phi)^R$ for all $w \in \Sig^*$. 
It follows that
\begin{align*}
w \in L^R 
&\iff
w^R \in L 
\iff
w^R \in K\phi^{-1}
\iff
(w^R)\phi \in K\\
&\iff
(w\phi)^R \in K
\iff
w\phi \in K^R
\iff 
w \in K^R\phi^{-1}.
\end{align*}
Thus $L^R = K^R\phi^{-1}$. 
Therefore, by Proposition \ref{prop:uniform}, reversal is uniform. \qed
\epf

The \e{cyclic shift} operation~\cite{JiOk08} is defined by $L^{\cyc} = \{uv : vu \in L\}$.
\bp
The cyclic shift operation $L \mapsto L^{\cyc}$ is uniform.
\ep
\bpf
Fix a 1-uniform morphism $\phi \co \Sig^* \ra \Gamma^*$ and suppose $L = K\phi^{-1}$. 
We want to show that $L^\cyc = K^\cyc \phi^{-1}$.

If $w \in L^\cyc$, we can write $w = uv$ for some $u,v \in \Sig^*$ such that $vu \in L$.
Since $L = K\phi^{-1}$, we have $(vu)\phi = (v\phi)(u\phi) \in K$.
Thus $(u\phi)(v\phi) = w\phi \in K^\cyc$, and 
it follows that $L^\cyc \subseteq K^\cyc\phi^{-1}$.

If $w \in K^\cyc\phi^{-1}$, then $w\phi \in K^\cyc$.
Thus we can write $w\phi = uv$ for some $u,v \in \Sig^*$ such that $vu \in K$.
Since $\phi$ is 1-uniform, $w$ has length $|u|+|v|$.
Write $w = xy$ where $|x| = |u|$ and $|y| = |v|$.
Then $(yx)\phi = (y\phi)(x\phi) = vu \in K$.
It follows that $yx \in L$, which implies $xy = w \in L^\cyc$.
Hence $L^\cyc = K^\cyc\phi^{-1}$.
By Proposition \ref{prop:uniform}, cyclic shift is uniform. \qed
\epf

\bp
\label{prop:star-uniform}
The star operation $L \mapsto L^*$ is uniform.
\ep
\bpf
Fix a 1-uniform morphism $\phi \co \Sig^* \ra \Gamma^*$ and suppose $L = K\phi^{-1}$. 
For a language $M$, let $t(M)$ be the set of all finite-length tuples of elements of $M$, and let $\psi_M \co t(M) \ra M^*$ be the map $(w_1,\dotsc,w_n)\psi_M = w_1\dotsb w_n$ (the empty tuple is sent to $\eps$).
We claim that $w\psi_L^{-1} \ne \emp \iff (w\phi)\psi_K^{-1} \ne \emp$. 
Indeed, if $(w_1,\dotsc,w_n) \in w\psi_L^{-1}$ then $(w_1\phi,\dotsc,w_n\phi) \in (w\phi)\psi_K^{-1}$.
Conversely, if we have $(x_1,\dotsc,x_n) \in (w\phi)\psi_K^{-1}$, then $w\phi = x_1\dotsb x_n$.
Since $\phi$ is 1-uniform, we can write $w = w_1\dotsb w_n$ with $|w_j| = |x_j|$ and $w_j\phi = x_j$ for $1 \le j \le n$.
Then since $x_j =  w_j\phi \in K \implies w_j \in K\phi^{-1} = L$, we have $(w_1,\dotsc,w_n) \in w\phi^{-1}_L$ as required.
It follows that:
\begin{align*}
w \in L^*
\iff 
w\psi^{-1}_L \ne \emp
\iff 
(w\phi)\psi^{-1}_K \ne \emp
\iff 
w\phi \in K^*
\iff 
w \in K^*\phi^{-1}.
\end{align*}
Thus $L^* = K^*\phi^{-1}$. 
Therefore, by Proposition \ref{prop:uniform}, star is uniform. \qed
\epf


An $m$-ary \e{boolean function} is a function $\beta \co \{0,1\}^m \ra \{0,1\}$.
Each $m$-ary boolean function defines a corresponding $m$-ary \e{boolean operation} on languages over $\Sig^*$, as follows.
For $L \subseteq \Sig^*$, let $\chi_L \co \Sig^* \ra \{0,1\}$ be the \e{characteristic function} of $L$: if $w \in L$ then $w\chi_L = 1$, and if $w \not\in L$ then $w\chi_L = 0$.
Then we define 
\[ (L_1,\dotsc,L_m)\beta = \{ w \in \Sig^* : (w\chi_{L_1},\dotsc,w\chi_{L_m})\beta = 1\}. \]
Examples of commonly used boolean operations on languages include union and intersection ($m$-ary for $m \ge 2$), difference and symmetric difference (binary), and complement (unary).
\bp
\label{prop:boolean}
Boolean operations on languages are uniform.
\ep
\bpf
Let $\beta$ be an $m$-ary boolean operation on languages.
Fix a 1-uniform morphism $\phi \co \Sig^* \ra \Gamma^*$ and suppose $L_j = K_j\phi^{-1}$ for $1 \le j \le m$.
We have
\begin{align*}
w \in (L_1,\dotsc,L_m)\beta
&\iff
(w\chi_{L_1},\dotsc,w\chi_{L_m})\beta = 1\\
&\iff
(w\chi_{K_1\phi^{-1}},\dotsc,w\chi_{K_m\phi^{-1}})\beta = 1\\
&\iff
(w\phi\chi_{K_1},\dotsc,w\phi\chi_{K_m})\beta = 1\\
&\iff
w\phi \in (K_1,\dotsc,K_m)\beta
\iff
w \in (K_1,\dotsc,K_m)\beta\phi^{-1}.
\end{align*}
Therefore, by Proposition \ref{prop:uniform}, $\beta$ is uniform. \qed
\epf
We have seen that binary concatenation is uniform, since concatenation belongs to the class of shuffles on trajectories. Next we give a direct proof that $m$-ary concatenation is uniform.
\bp
The $m$-ary concatenation operation $(L_1,\dotsc,L_m)\!\mapsto\!L_1\dotsb L_m$ is uniform.
\ep
\bpf
Fix a 1-uniform morphism $\phi \co \Sig^* \ra \Gamma^*$ and suppose $L_j = K_j\phi^{-1}$ for $1 \le j \le m$.
Define $\psi_L \co L_1 \times \dotsb \times L_m \ra L_1\dotsb L_m$ by $(w_1,\dotsc,w_m)\psi_L = w_1\dotsb w_m$
and similarly define $\psi_K \co K_1 \times \dotsb \times K_m \ra K_1\dotsb K_m$.
Using similar arguments to the proof of Proposition \ref{prop:star-uniform}, we can show that $w\psi_L^{-1} \ne \emp \iff (w\phi)\psi_K^{-1} \ne \emp$. Thus:
\begin{align*}
w \in L_1\dotsb L_m
\iff 
w\psi^{-1}_L \ne \emp
\iff 
(w\phi)\psi^{-1}_K \ne \emp
\iff 
w\phi \in K_1 \dotsb K_m.
\end{align*}
Therefore, by Proposition \ref{prop:uniform}, $m$-ary concatenation is uniform. \qed
\epf
Next, we show that the class of uniform operations is closed under composition. It is easy to see that this holds for unary uniform operations:
if $\Phi$ and $\Phi'$ are uniform and $L = K\phi^{-1}$ for a 1-uniform morphism $\phi$, then $L\Phi = K\Phi\phi^{-1}$ and subsequently $(L\Phi)\Phi' = (K\Phi)\Phi'\phi^{-1}$.
The general case is not much harder; the only difficulty is in dealing with the notation.
\bp
\label{prop:comp}
Let $\Phi$ be an $m$-ary uniform operation, and let $\Phi_1,\dotsc,\Phi_m$ be uniform operations where $\Phi_j$ has arity $n_j$. Set $N_j = n_1 + \dotsb +n_j$ for $1 \le j \le m$, and consider the operation of arity $N_m$ that maps $(L_1,\dotsc,L_{N_m})$ to
\[
( 
(L_1,\dotsc,L_{N_1})\Phi_1,
(L_{N_1+1},\dotsc,L_{N_2})\Phi_2,
\dotsc
(L_{N_{m-1}+1},\dotsc,L_{N_m})\Phi_m
)\Phi.
\]
This operation, which we denote by $\Phi'$, is uniform.
\ep
\bpf
Fix a 1-uniform morphism $\phi \co \Sig^* \ra \Gamma^*$ and suppose $L_j = K_j\phi^{-1}$ for $1 \le j \le N_m$.
By Proposition \ref{prop:uniform}, it suffices to show that 
$(L_1,\dotsc,L_{N_m})\Phi' = (K_1,\dotsc,K_{N_m})\Phi'\phi^{-1}$. 
Let $N_0 = 0$; then by the uniformity of $\Phi_j$, for $1 \le j \le m$ we have 
\[ 
(L_{N_{j-1}+1},\dotsc,L_{N_j})\Phi_j =
(K_{N_{j-1}+1},\dotsc,K_{N_j})\Phi_j\phi^{-1}. 
\]
Set 
$M_j = (L_{N_{j-1}+1},\dotsc,L_{N_j})\Phi_j$ 
and
$M'_j = (K_{N_{j-1}+1},\dotsc,K_{N_j})\Phi_j$.
Then $M_j = M'_j\phi^{-1}$ for $1 \le j \le m$.
By the uniformity of $\Phi$, we have
{\small
\[ 
(L_1,\dotsc,L_{N_m})\Phi' =
(M_1,\dotsc,M_m)\Phi = 
(M'_1,\dotsc,M'_m)\Phi\phi^{-1} =
(K_1,\dotsc,K_{N_m})\Phi'\phi^{-1},
\]
}
as required. \qed
\epf
This shows that all ``combined operations'' formed by compositions of the uniform operations we have seen so far are also uniform.

The following ``substitution lemma'' can also be used to construct new uniform operations from known ones.
\bl
\label{lem:sub}
Let $\Phi$ be a $k$-ary operation. Fix $m \ge 1$ and $i_1,\dotsc,i_k \in \{1,\dotsc,m\}$.
Then the operation $\Phi'$ defined by $(L_1,\dotsc,L_m) \mapsto (L_{i_1},\dotsc,L_{i_k})\Phi$ is uniform.
\el
\bpf
Fix a 1-uniform morphism $\phi \co \Sig^* \ra \Gamma^*$ and suppose $L_j = K_j\phi^{-1}$ for $1 \le j \le m$.
Then by the uniformity of $\Phi$, we have
\[
(L_1,\dotsc,L_m)\Phi' = (L_{i_1},\dotsc,L_{i_k})\Phi = (K_{i_1},\dotsc,K_{i_k})\Phi\phi^{-1}
= (K_1,\dotsc,K_m)\Phi'\phi^{-1}. \]
Therefore, by Proposition \ref{prop:uniform}, the operation $\Phi'$ is uniform. \qed
\epf
As an example, we show that the power operation is uniform. Define $L^0 =\{\eps\}$ and for $n \ge 1$, set $L^n = L^{n-1}L$.
\bp
For $n \ge 0$, the power operation $L \mapsto L^n$ is uniform.
\ep
\bpf
For $n \ge 2$, in Lemma \ref{lem:sub}, let $\Phi$ be the $n$-ary concatenation operation, set $m = 1$ and set $i_1,\dotsc,i_n = 1$. 
For $n = 1$, it is immediate that $L \mapsto L$ is uniform.
For $n = 0$, see Proposition \ref{prop:constant} below.
\qed
\epf
Another example is the anti-concatenation operation $(L,L') \mapsto L'L$. This belongs to the class of shuffles on trajectories, so we already know that it is uniform, but an alternate proof could be given using Lemma \ref{lem:sub}: let $\Phi$ be binary concatenation, set $m = 2$, set $i_1 = 2$ and set $i_2 = 1$.

Next we consider some operations which depend only on the alphabet of the input languages. These are not interesting from a state complexity perspective, but can be used to construct interesting combined operations.
\bp
\label{prop:constant}
Let $S \subseteq \bN$. The operation $(L_1,\dotsc,L_m) \mapsto \bigcup_{n \in S} \Sig^n$, where $\Sig$ is the common alphabet of the inputs, is uniform. In particular, the following operations are uniform for all arities $m$:
\be
\item
$(L_1,\dotsc,L_m) \mapsto \emp$.
\item
$(L_1,\dotsc,L_m) \mapsto \{\eps\}$.
\item
$(L_1,\dotsc,L_m) \mapsto \Sig^*$.
\item
$(L_1,\dotsc,L_m) \mapsto \Sig^+$.
\ee
\ep
\bpf
Fix a 1-uniform morphism $\phi \co \Sig^* \ra \Gamma^*$ and suppose $L_j = K_j\phi^{-1}$ for $1 \le j \le m$.
We claim that $\Sig^n = \Gamma^n\phi^{-1}$ for all $n \ge 0$.
Indeed, take a word $w \in \Sig^n$; then $w\phi$ is in $\Gamma^n$ by 1-uniformity, and so $w \in \Gamma^n\phi^{-1}$. Conversely, if $w \in \Gamma^n\phi^{-1} = \{ x \in \Sig^* : x\phi \in \Gamma^n\}$ then certainly $w \in \Sig^n$. It follows that:
\[ 
\bigcup_{n \in S} \Sig^n = 
\bigcup_{n \in S} \Gamma^n\phi^{-1} =
\left(\bigcup_{n \in S} \Gamma^n\right)\phi^{-1}. 
\]
By Proposition \ref{prop:uniform}, operations of this type are uniform. \qed
\epf
For a language $L$ over $\Sig$, the \e{right ideal} generated by $L$ is $\Sig^*L$, the \e{left ideal} generated by $L$ is $L\Sig^*$, the \e{two-sided ideal} generated by $L$ is $\Sig^*L\Sig^*$, and the \e{all-sided ideal} generated by $L$ is $L \shu \Sig^*$, where $\shu$ is (ordinary) shuffle. 
By combining our earlier results, we can show that the operations which map $L$ to one of the ideals it generates are uniform.
For example, let $\Phi$ be ternary concatenation, let $\Phi_1$ and $\Phi_3$ be $L \mapsto \Sig^*$, and let $\Phi_2$ be $L \mapsto L$. Then the operation $(L_1,L_2,L_3) \mapsto (L_1\Phi_1,L_2\Phi_2,L_3\Phi_3)\Phi$ is uniform by closure under composition. Then by substitution, $L \mapsto (L\Phi_1,L\Phi_2,L\Phi_3)\Phi = \Sig^*L\Sig^*$ is uniform.

In summary, we have proved that the following operations are uniform: reversal, cyclic shift, star, power, $m$-ary concatenation, $m$-ary boolean operations (including union, intersection, difference, symmetric difference and complement), shuffles on trajectories (including shuffle, literal shuffle, balanced literal shuffle, insertion, balanced insertion, and anti-concatenation), and the ``alphabet-dependent'' operations of Proposition \ref{prop:constant}. We also proved that the class of uniform operations is closed under composition, meaning that \e{all combined operations} formed by composing the aforementioned operations are uniform, such as ``star-complement-star'' or ``star of union''. Additionally, we proved a substitution lemma that gives another way to construct new uniform operations from old, such as the ``ideal generated by'' operations.

\subsection{Non-Uniform Operations}
First we remark that \e{constant operations}, which output a fixed language regardless of the input, are not in general uniform.
One issue is that our theoretical framework assumes that all regular operations are \e{alphabet-preserving}, so we cannot even define true ``constant operations'' that take arbitrary regular languages as inputs; we must restrict the inputs to have the same alphabet as the constant output language.
The more fundamental problem is that constant operations need not behave uniformly with respect to transformations. For example, let $\Psi$ be a constant DFA operation, and suppose that in DFA $\cA$, the letter $a$ induces transformation $t$, and in DFA $\cB$, the letter $b$ also induces transformation $t$. If $\Psi$ is uniform, then the transformation induced by $a$ in $\cA\Psi$ will be the same as the transformation induced by $b$ in $\cB\Psi$. But the constant operation $\Psi$ could produce a DFA in which $a$ and $b$ induce different transformations, violating uniformity. The only way to ensure uniformity is if $\Psi$ produces a DFA in which every letter induces the same transformation; if we enforce this condition, we essentially obtain the alphabet-dependent operations of Proposition \ref{prop:constant}.

Our first example of an interesting non-uniform operation is the following:
\[ \frac{1}{2}L = \{ x \in \Sig^* : xy \in L, |x| = |y|\}. \]
This ``half'' operation is an example of a \e{proportional removal}; the state complexity of proportional removals was studied by Domaratzki~\cite{Dom02}. 
We could prove that this operation is not uniform directly from the definition, or using the language-theoretic characterization, but instead we will show something even stronger: the OLPA approach does not maximize the state complexity of this operation.

If the OLPA approach worked for this operation, then by Lemma \ref{lem:olpa}, the state complexity of the operation would be maximized by a language of the form $\frac{1}{2}T_Q(i,F)$ for some state configuration $(Q,i,F)$. However, it is not hard to see that if $F$ is non-empty, then $\frac{1}{2}T_Q(i,F)$ is either $T_Q^*$ or $T_Q^* \setminus \{\eps\}$, depending on whether $i \in F$.
Indeed, let $w$ be a non-empty word in $T_Q^*$. We have $iw = q$ for some $q \in Q$. Let $t$ be a transformation that sends $q$ into $F$. Then $wt\id_Q^{|w|-1}$ maps $i$ into $F$, and so this word is in the language $T_Q(i,F)$. But $w$ is exactly half the length of this word, so $w \in \frac{1}{2}T_Q(i,F)$. This means that $\stc(\frac{1}{2}T_Q(i,F)) \le 2$; but Domaratzki~\cite{Dom02} shows that there are languages $L$ of state complexity $n$ such that $\stc(\frac{1}{2}L) = n$.
A similar argument shows that OLPA approach fails for many other proportional removal operations as well, although we have not tried to characterize the proportional removals for which the approach fails.

Next we consider \e{deletions along trajectories}~\cite{Dom04,HKS16}, a class of operations which includes left quotient, right quotient, deletion, scattered deletion, bi-polar deletion, and $k$-deletion~\cite[p.\ 296]{Dom04}. We will show that the left quotient operation and the deletion operation are not uniform.
We have not investigated uniformity for other deletions along trajectories.

The case of left quotient is interesting, because the OLPA approach actually works for this operation despite its non-uniformity. The left quotient of $L$ by $L'$ is 
$L'\lquo L = \{ x \in \Sig^* : wx \in L \text{ for some $w \in L'$}\}$.
This operation satisfies a weak version of the language-theoretic characterization of uniformity:

\e{For all $1$-uniform morphisms $\phi \co \Sig^* \ra \Gamma^*$, if $L_j = K_j\phi^{-1}$ and $L_j \ne \emp$ for $1 \le j \le 2$, then ($L_2\lquo L_1)\Phi = (K_2 \lquo K_1)\Phi\phi^{-1}$.}

Because empty languages are excluded here, the OLPA approach would fail if maximizing the state complexity in certain cases required the use of empty languages. But this does not happen for left quotient.

To see that left quotient is not uniform, let $\Sig = \{a,b\}$ and define $\phi \co \Sig^* \ra \Sig^*$ by $a\phi = b\phi = b$. 
Then define $K_1 = \{ab\}$, $K_2 = \{a\}$, $L_1 = K_1\phi^{-1}$, and $L_2 = K_2\phi^{-1}$.
If left quotient was uniform, we would have $L_2 \lquo L_1 = (K_2 \lquo K_1)\phi^{-1}$.
But $L_1 = L_2 = \emp$, and so $L_2 \lquo L_1 = \emp$.
Meanwhile, $(K_2 \lquo K_1)\phi^{-1} = (\{a\}\lquo \{ab\})\phi^{-1} = \{b\}\phi^{-1} = \{a,b\}$.

The \e{deletion} of $L'$ from $L$ is
$L \del L' = \{ xz \in \Sig^* : xyz \in L \text{ for some $y \in L'$}\}$.
We will show that the OLPA approach fails for this operation.

If the OLPA approach worked, the state complexity would be maximized by some pair of OLPA witnesses.
Consider the language 
$\bT_1(i_1,F_1) \del \bT_2(i_2,F_2)$ 
where $\bT = \cT_{Q_1} \times \cT_{Q_2}$ for some finite sets $Q_1$ and $Q_2$.
We claim that
$\bT_1(i_1,F_1) \del \bT_2(i_2,F_2) = \bT^*$, which has state complexity one.

Indeed, fix a word $w \in \bT^*$.
Write $w = (t_{1,1},t_{2,1}) \dotsb (t_{1,k},t_{2,k})$.
Let $w_1 = t_{1,1} \dotsb t_{1,k}$,
set $q_1 = i_1w_1$, and choose a transformation $t_1 \co Q_1 \ra Q_1$ that sends $q_1$ into $F_1$.
Next, choose a transformation $t_2 \co Q_2 \ra Q_2$ that sends $i_2$ into $F_2$.
Then $i_1 w_1 t_1 \in F_1$, so $w(t_1,t_2) \in \bT_1(i_1,F_1)$.
However, $i_2 t_2 \in F_2$, so $(t_1,t_2) \in \bT_2(i_2,F_2)$.
It follows $w \in \bT_1(i_1,F_1) \del \bT_2(i_2,F_2)$ since it can be obtained by deleting a word in $\bT_2(i_2,F_2)$ from a word in $\bT_1(i_1,F_1)$.

This shows that using OLPA witnesses for deletion only produces languages of state complexity one. However, Han, Ko and Salomaa~\cite{HKS16} proved that if $L$ has state complexity $n$, then $n2^{n-1}$ is a tight upper bound on the state complexity of $L \del L'$. Hence the state complexity of deletion is not maximized by the OLPA approach.

It is interesting that our main examples of operations for which the OLPA approach fails involve the idea of ``deletion'' in some sense.

\section{Proofs using the OLPA Approach}
In the introduction, we used the OLPA approach to give a simple proof of the worst-case state complexity of reversal.
We give two additional examples of proofs using the OLPA approach in this section.
First we consider the star operation.
\bp
Let $L$ be a regular language recognized by $\cD = (Q,\Sig,T,i,F)$, where $|Q| = n$ and $|F| = k$. Suppose $1 \le |F| \le n-1$. If $F = \{i\}$ then $L = L^*$, and otherwise we have the following tight upper bounds on $\stc(L^*)$:
\[ \stc(L^*) \le \begin{cases} 
(2^{n-k}-1)+(2^{n-1}-2^{n-k-1})+1, & \text{if $i \not\in F$;}\\
(2^{n-k}-1)+(2^{n-1}), & \text{if $i \in F$ and $|F| \ge 2$.}\\
\end{cases}
\]
\ep
\bpf
By Theorem \ref{thm:olpa}, it suffices to just compute the state complexity of $L^*$ for
$L \in \{ \cT_{n}(1,F_{n,k,j}) : 0 \le j \le 1, 0+j \le k \le n-1+j\}$.
Given $\cD$, an FA for $L^*$ is $\cA = (Q \cup \{s\},\Sig,T',s,F \cup \{s\})$ where $T' = T \cup \{(f,a,iT_a) : f \in F \cup\{s\}\}$.
It is easy to see that if $F = \{i\}$, then $\cA$ recognizes $L$.
Henceforth assume $F \ne \{i\}$.

We show each non-empty set $S \subseteq Q$ is reachable by induction on $|S|$.
From $\{s\}$ we reach $\{q\}$ for each $q \in Q$ by a transformation that sends $i$ to $q$. 
Now suppose $|S| \ge 2$ and smaller sets are reachable.
Choose a set $X$ of size $|S|-1$ which contains a final state but does not contain $i$; this is possible since $F \ne \{i\}$.
Fix $q \in S$ and choose a transformation $t$ that maps $X$ onto $S \setminus \{q\}$ and $i$ to $q$; then $t$ sends $X$ to $S$.
Thus all non-empty subsets of $Q$ are reachable.

We show that sets in the following collection are pairwise distinguishable:
\[ \{S \subseteq Q: S \ne \emp \text{ and } S \cap F = \emp\} \cup \{ S \subseteq Q: i \in S \text{ and } S \cap F \ne \emp \}. \]
If a non-empty set $S$ is not in this collection, it contains a final state but does not contain $i$, and the set $S$ is indistinguishable from $S \cup \{i\}$.
To distinguish distinct sets $S$ and $X$ in this collection, choose an element $q$ which appears (without loss of generality) in $S$ but not in $X$, and apply a transformation $t$ which maps $q$ into $F$ and $Q \setminus \{q\}$ into $Q \setminus F$.
Note also that if $i \not\in F$ then $\{s\}$ is distinguishable from all states in this collection, but if $i \in F$ then $\{s\}$ and $\{i\}$ are indistinguishable. 
Elementary counting arguments then yield the stated bounds.
%
\qed
\epf

For our other example, we consider boolean operations, defined in Section \ref{sec:examples} (see the discussion before Proposition \ref{prop:boolean}). In this case the proof is complicated, but the result is very general, and we believe it would be considerably more difficult to prove without the OLPA approach or a similar construction.

It is a bit tricky to state a tight upper bound for the worst-case state complexity of an arbitrary $m$-ary boolean operation, because the operation's result might not depend on all of its arguments. For example, if the inputs to a binary boolean operation have state complexity $n_1$ and $n_2$ respectively, the worst-case state complexity can be $1$, $n_1$, $n_2$, or $n_1n_2$, depending on which arguments (if any) are relevant to the result. 

To state our upper bound, we introduce some notation. Given an $m$-ary boolean function $\beta \co \{0,1\}^m \ra \{0,1\}$, we define functions $\beta_j \co \bN \ra \bN$ for $1 \le j \le m$ as follows. 
If there exist two binary $m$-tuples $(b_1,\dotsc,b_m)$ and $(b'_1,\dotsc,b'_m)$ which differ only in the $j$-th bit (that is, $b_j \ne b'_j$ and $b_i = b'_i$ for all $i \ne j$) such that $(b_1,\dotsc,b_m)\beta \ne (b'_1,\dotsc,b'_m)\beta$, then we define $n\beta_j = n$ for all $n \in \bN$. 
Otherwise, it must be the case that for all binary $m$-tuples, flipping the $j$-th bit does not change the result of $\beta$; in this case we define $n\beta_j = 1$ for all $n \in \bN$. 
If $\beta_j$ is the identity map, we say that $\beta$ \e{depends on argument $j$}, and if $\beta_j$ is the constant function sending everything to $1$, we say that $\beta$ \e{does not depend on argument $j$}.
\bp
Let $\beta$ be an $m$-ary boolean operation. Let $(L_1,\dotsc,L_m)$ be regular languages, where $L_j$ is recognized by $(Q_j,\Sig,T_j,i_j,F_j)$ for $1 \le j \le m$. Set $n_j = |Q_j|$. Then $\stc((L_1,\dotsc,L_m)\beta) \le (n_1\beta_1) \dotsb (n_m\beta_m)$ and this bound is tight. 
\ep
\bpf
Recall the usual direct product construction for boolean operations:
$\cD = (Q,\Sig,T,(i_1,\dotsc,i_m),F)$ where the state set is $Q = Q_1 \times \dotsb \times Q_m$,
the transition set is $T = \{ ((q_1,\dotsc,q_m),a,(q_1(T_1)_a,\dotsc,q_m(T_m)_a)) : (q_1,\dotsc,q_m) \in Q, a \in \Sig\}$
and the final state set is
$F = \{ (q_1,\dotsc,q_m) \in Q : (q_1\chi_{F_1},\dotsc,q_m\chi_{F_m}) = 1\}$.
This construction gives an upper bound of $n_1\dotsb n_m$. To get a tighter bound, we must consider distinguishability.
Consider the following set of states:
\[ \{ (q_1,\dotsc,q_m)  \in Q : \text{$q_j = i_j$ whenever $\beta$ does not depend on argument $j$}\}. \]
There are precisely $(n_1\beta_1)\dotsb (n_m\beta_m)$ states in this set, and we claim that every state lying outside this set is indistinguishable from a state within the set.
To see this, fix a state $(q_1,\dotsc,q_m)$ which is not in the above set. Then there exists $j$ such that $q_j \ne i_j$ and $\beta$ does not depend on argument $j$. We claim state $(q_1,\dotsc,q_m)$ is indistinguishable from $(q_1,\dotsc,q_{j-1},i_j,q_{j+1},\dotsc,q_m)$. Indeed, if two states differ only in component $j$, and $\beta$ does not depend on argument $j$, then either both states are final or both states are non-final. Also, starting from a pair of states which differ only in component $j$, we can only reach other pairs that differ only in component $j$. Thus there is no way to distinguish these states.

Now we show that the upper bound is tight. Our witnesses will be OLPA witnesses $L_j = \bT_j(1,F_j)$,
where $\bT = \cT_{n_1} \times \dotsb \times \cT_{n_m}$ and $F_j = \{1\}$ for $1 \le j \le m$ (it does not really matter what we choose for $F_j$, as long as for $n_j \ge 2$ it is a proper non-empty subset of $\{1,\dotsc,n_j\})$.

The initial state of the direct product DFA is $(1,1,\dotsc,1)$. For each state $(q_1,\dotsc,q_m) \in Q$, let $t_j$ be a transformation that sends $1$ to $q_j$. Then the letter $(t_1,\dotsc,t_m) \in \bT$ sends the initial state to $(q_1,\dotsc,q_m)$; thus all states are reachable. 

Next we show that all pairs of states in $\{ (q_1\beta_1,\dotsc,q_m\beta_m) : (q_1,\dotsc,q_m) \in Q \}$,
which has size $(n_1\beta_1)\dotsb(n_m\beta_m)$, are distinguishable.
Suppose we have two distinct states 
$(q_1\beta_1,\dotsc,q_m\beta_m)$
and
$(q'_1\beta_1,\dotsc,q'_m\beta_m)$.
Since they are distinct, they must differ in some component $j$, and for this $j$ we must have $q_j\beta_j = q_j$ and $q'_j\beta_j = q'_j$.
Hence there exist two binary $m$-tuples $(b_1,\dotsc,b_m)$ and $(b'_1,\dotsc,b'_m)$ which differ only in component $j$ such that $(b_1,\dotsc,b_m)\beta \ne (b'_1,\dotsc,b'_m)\beta$.
Assume without loss of generality that $(b_1,\dotsc,b_m)\beta = 1$ and $(b'_1,\dotsc,b'_m)\beta = 0$.

Now, choose a tuple of transformations $(t_1,\dotsc,t_m) \in \bT$ as follows:
\bi
\item
Choose $t_j$ so that $q_jt_j\chi_{F_j} = b_j$ and $q'_jt_j\chi_{F_j} = b'_j$.
\item
For $i \ne j$,
if $\beta$ depends on argument $i$, choose $t_i$ so that $(q_it_i)\chi_{F_i} = b_i = b'_i$. 
If $\beta$ does not depend on argument $i$, let $t_i$ be the identity map.
\ei
Now, we claim that $(q_1\beta_1,\dotsc,q_m\beta_m)(t_1,\dotsc,t_m)$ is a final state.
To determine whether the reached state $(q_1\beta_1t_1,\dotsc,q_m\beta_mt_m)$ is final, we look at the binary $m$-tuple
$(q_1\beta_1t_1\chi_{F_1},\dotsc,q_m\beta_mt_m\chi_{F_m})$.
If $\beta$ depends on argument $i$ (including the case $i = j$) then we have $q_i\beta_i t_i\chi_{F_i} = b_i$.
If $\beta$ does not depend on argument $i$, then $q_i\beta_i = 1$, transformation $t_i$ is the identity map, and $F_i = \{1\}$, so we have $q_i\beta_it_i\chi_{F_i} = 1$.
Thus $(q_1\beta_1t_1\chi_{F_1},\dotsc,q_m\beta_mt_m\chi_{F_m}) = (\wt{b_1},\dotsc,\wt{b_m})$, where $\wt{b_i}$ is $b_i$ if $\beta$ depends on argument $i$, and $\wt{b_i}$ is $1$ otherwise.
Now, we know that if $\beta$ does not depend on argument $i$, then flipping the $i$-th bit in a binary $m$-tuple will not change the result of applying $\beta$ to that $m$-tuple. So by flipping bits if necessary, we see that
\[ (\wt{b_1},\dotsc,\wt{b_m})\beta = (b_1,\dotsc,b_m)\beta = 1. \]
Hence $(q_1\beta_1,\dotsc,q_m\beta_m)(t_1,\dotsc,t_m)$ is a final state.

On the other hand, consider the state $(q'_1\beta_1,\dotsc,q'_m\beta_m)(t_1,\dotsc,t_m)$.
For this state we have $(q'_1\beta_1t_1\chi_{F_1},\dotsc,q'_m\beta_mt_m\chi_{F_m})
= (\wt{b'_1},\dotsc,\wt{b'_m})$, where $\wt{b'_i}$ is $b'_i$ if $\beta$ depends on argument $i$, and $\wt{b'_i}$ is $1$ otherwise. Thus by the same bit-flipping argument, we have
\[ (\wt{b'_1},\dotsc,\wt{b'_m})\beta = (b'_1,\dotsc,b'_m)\beta = 0. \]
Thus this state is not final, and we have distinguished the two states. \qed
\epf

\section{Conclusions}
The ``one letter per action'' (OLPA) approach gives an easy way to find witnesses that maximize the state complexity of many regular operations, at the expense of requiring large alphabets.
We defined a class of ``uniform'' regular operations for which the OLPA approach provably works.
This class contains many common operations and is also closed under composition.
We hope this paper will spark interest in and further study of the OLPA approach.

We list a few open questions that we find interesting.
\bi
\item
To what extent does the OLPA approach work in \e{subclasses} of the regular languages? It seems it will work for some ``nice'' subclasses but not for others.
\item
Can the OLPA approach be generalized to other state complexity measures, like incomplete state complexity~\cite{MMR15} or unrestricted state complexity~\cite{Brz16}?
\item
Can we find a larger class than the class of uniform operations for which the OLPA approach provably works, without sacrificing the nice property of closure under composition?
\item
How do we maximize the state complexity of proportional removals like $\frac{1}{2}L$? Domaratzki's work~\cite{Dom02} does not completely solve this problem. If we find a way to maximize their state complexity, can it be generalized to other operations for which the OLPA approach fails?
\ei

\noindent
{\bf Acknowledgements.} I thank 
Jason Bell, Janusz Brzozowski, and the referees of the DLT 2018 version of this paper
for their careful proofreading and helpful comments.
I thank Lukas Fleischer for pointing me to some important references I overlooked, which allowed me to give a much more complete history of the ideas presented in this paper.

\bibliographystyle{splncs03}
\bibliography{theory}
\end{document}